%% file: correia.tex
\documentclass[multphys,vecphys]{svmult}

\def\farcs{\hbox{$.\!\!^{\prime\prime}$}}
\def\fdg{\hbox{$.\!\!^\circ$}}
\def\la{\mathrel{\mathchoice {\vcenter{\offinterlineskip\halign{\hfil
$\displaystyle##$\hfil\cr<\cr\sim\cr}}}
{\vcenter{\offinterlineskip\halign{\hfil$\textstyle##$\hfil\cr
<\cr\sim\cr}}}
{\vcenter{\offinterlineskip\halign{\hfil$\scriptstyle##$\hfil\cr
<\cr\sim\cr}}}
{\vcenter{\offinterlineskip\halign{\hfil$\scriptscriptstyle##$\hfil\cr
<\cr\sim\cr}}}}}

\usepackage{makeidx}         
\usepackage{graphicx}        
\usepackage{multicol}        
\usepackage[bottom]{footmisc}

\makeindex             

\begin{document}

\title*{First evidence for a spatially resolved disk structure around the Herbig Ae star R CrA}
\titlerunning{Resolved mid-infrared disk structure around R CrA}
\author{S. Correia\inst{1}
          \and
          R. K\"ohler\inst{2}
          \and
          G. Meeus\inst{1}
          \and 
          H. Zinnecker\inst{1}
          }

\institute{Astrophysikalisches Institut Potsdam, An der Sternwarte 16, D-14482 Potsdam, Germany
              \texttt{scorreia@aip.de, hzinnecker@aip.de, gwen@aip.de}
         \and
             Sterrewacht Leiden, Niels Bohrweg 2, NL-2333 CA Leiden, The Netherlands
             \texttt{koehler@strw.leidenuniv.nl}
             }

%
\maketitle

\begin{abstract}
We present mid-infrared interferometric observations of the Herbig Ae star R CrA obtained with MIDI at the VLTI using several projected baselines 
in the UT2-UT3 configuration. The observations show resolved circumstellar emission in the wavelength range 8-13\,$\mu$m on a $\sim$\,6\,AU scale 
with a non-symmetric intensity distribution, providing support for an inclined disk geometry. Visibilities are best fitted using a uniform ring model 
with outer radius in the range 6-10 AU, in the wavelength range 8-13 micron. The inclination of the ring with respect to the plane 
of the sky is found to be $\sim$\,45$^\circ$, consistent with the 40$^\circ$ suggested from near-infrared imaging polarimetry (Clark et al. 2000, MNRAS, 319, 337). 

\end{abstract}

\section{Introduction}
\label{sec:intro}


The presence of circumstellar disks around the intermediate mass (M$\la$5M$_\odot$) Herbig Ae stars is supported by a large body of observational 
evidence\,\cite{Natta _etal_2000}\cite{Hillenbrand_etal_1992}. More specifically, while the observed spectral energy distribution (SED) of such stars 
can be explained by both a disk-like distribution of material (e.g.\,\cite{Dullemond_etal_2001}) and other geometries like envelopes (e.g.\,\cite{Hartmann_etal_1993}), 
clear evidence for circumstellar disks comes from resolved flattened structures observed by interferometry at millimeter, near-IR and recently also mid-IR 
wavelengths (e.g.\,\cite{Mannings_Sargent_1997}\cite{Eisner_etal_2004}\cite{Leinert_etal_2004}).
 
R CrA is a bright (100 L$_\odot$) young Herbig A5e star, located at the center of a small cluster (The Coronet\,\cite{Taylor_Storey_1984}\cite{Wilking_etal_1997}) 
at 130~pc\,\cite{De_Zeeus_etal_1999}. Several characteristics indicate the presence of a circumstellar disk around R CrA\,: a flat mid-IR to far-IR/mm 
SED\,\footnote{although most of the mm-excess is actually from the nearby embedded infrared source IRS7\,\cite{Choi_Tatematsu_2004} and source confusion in the large IRAS beams might be an issue.} (\cite{Hillenbrand_etal_1992}), a broad silicate emission feature\,\cite{Acke_Van_Ancker_2004}, a UX Ori Type\,\cite{Dullemond_etal_2003}, 
a high degree (8\%) of optical linear polarisation\,\cite{Bastien_1987} , the possible association with a extended molecular outflow\,\cite{Levreault_1988},\cite{Anderson_etal_1997} 
as well as with several Herbig-Haro systems\,\cite{Wang_2004}, and a near-infrared reflection nebulosity whose resolved spatial polarization is consistent with a bipolar outflow being truncated by an evacuated spherical cavity\,\cite{Clark_etal_2000}.


\begin{figure}
\centering
\begin{tabular}{cc}
\includegraphics[height=4.5cm]{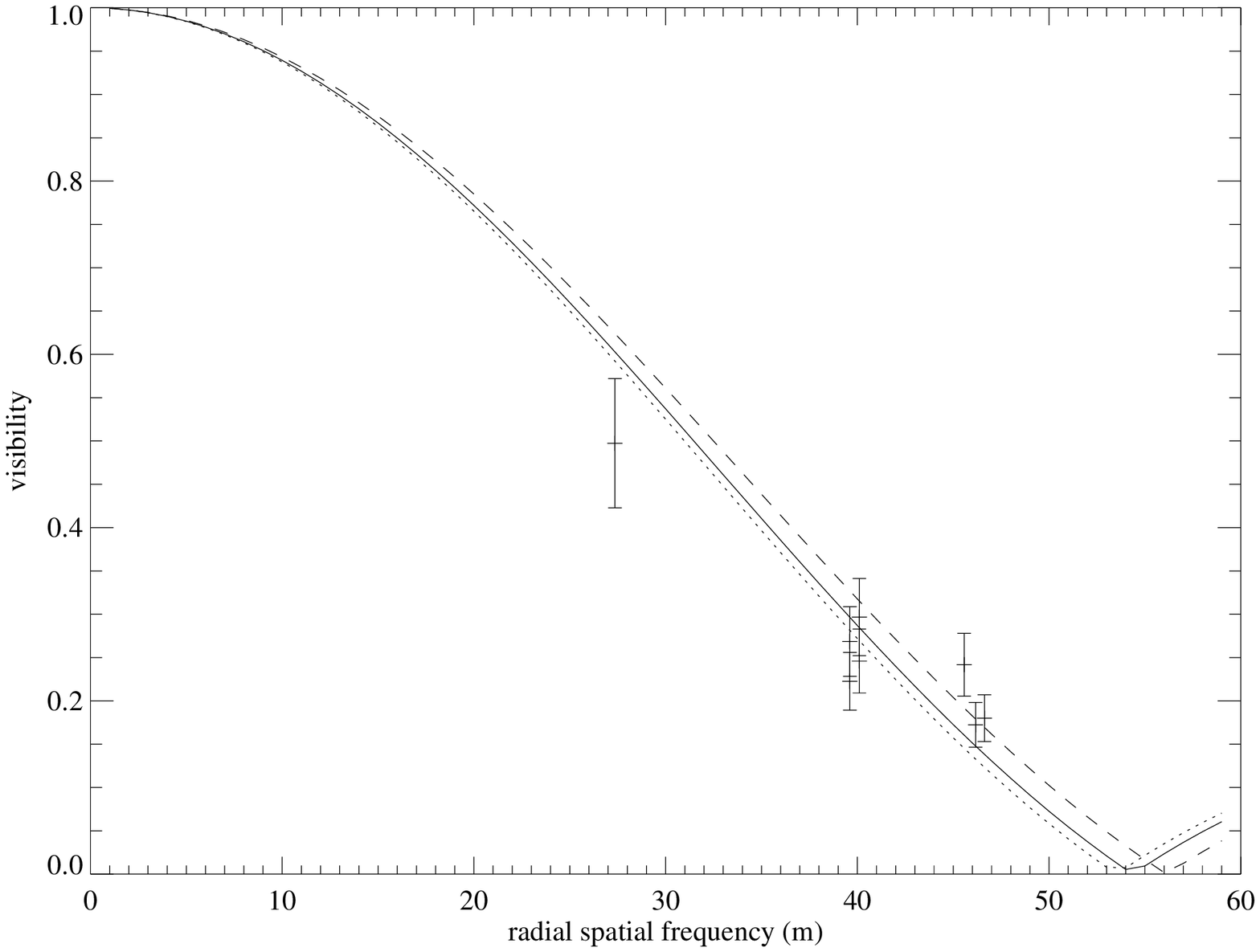} &
\includegraphics[height=5.5cm]{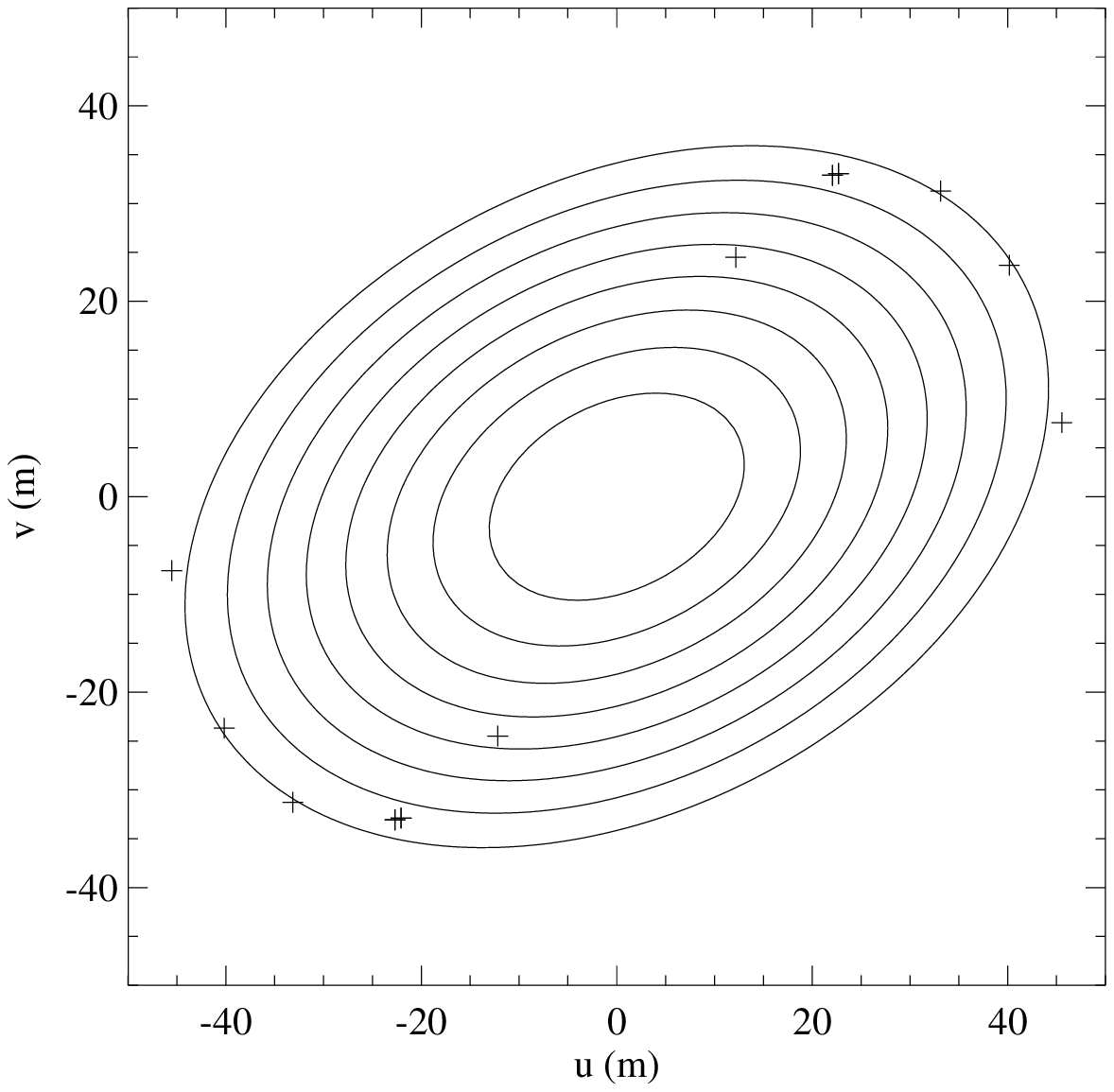}
\end{tabular}
\caption{Left panel\,: Wavelength averaged visibilities as a function of {\it uv}-radius and best-fit face-on uniform ring brightness distribution (R$_{out}$=6.3\,AU$^{+0.1}_{-0.2}$). 
Dashed and dotted lines are respectively the inner and outer boundary values of R$_{out}$.  
Right panel\,:Contour plot of the best-fit inclined uniform ring model whose parameters are listed in Table\,\ref{tab: bestfit_disks_par}.
The contour levels are for visibilities of 0.1, 0.2, 0.3, 0.4, 0.5, 0.6, 0.7, 0.8, 0.9. East is up, north is right.
}
\label{fig: face-on-incl-ring}      
\end{figure}

\section{Observations and Calibration}
\label{sec: obs_cal}

R CrA has been observed with MIDI at the VLTI between 2004 July 8 and 30. We obtained 6 sets of spectrally dispersed (Prism, R=30) visibilities 
in the spectral range 8-13\,$\mu$m with the 47\,m long northeast UT2-UT3 baseline. Projected baseline lengths and PAs are 27.4\,m, 46.6\,m, 
46.2\,m, 45.6\,m, 40.1\,m, 39.6\,m and 63$\fdg$6, 30$\fdg$5, 9$\fdg$4, 43$\fdg$4, 55$\fdg$6, 56$\fdg$1, respectively. 
Each data set has been reduced with the MIA package\,\cite{Koehler_MIA}.
The raw visibilities were calibrated by observing the source HD 173484, whose adopted angular size is 3.35\,$\pm$0.31\,mas (MIDI calibrator list). 
We evaluated the error on the visibilities by comparing the instrumental visibility obtained with different calibrators observed during the same night. 
This leads to a typical relative uncertainty of 15\%, a value we adopt in the following. 
The acquisition images from both telescopes show that R CrA is unresolved at 8.7\,$\mu$m, i.e. at a resolution limit of $\lambda/D$=0$\farcs$27, 
corresponding to $\sim$\,30\,AU at 130\,pc.


\section{Evidence for an inclined disk structure}
\label{sec: results}

\subsection{First approach}
\label{sec:first_approach}

We compared the observed visibilities with those derived from a simple geometric disk model chosen to be a uniform ring brightness distribution. 
The inner radius was fixed to a dust sublimation radius of 0.3\,AU, typical for a Herbig AeBe star of that luminosity\,\cite{Monnier_etal_2005}. 
In a first step, only the wavelength averaged visibilities were fitted. We will see later that this rough approximation can be justified by the relative 
flatness of the visibilities observed on R CrA in the 8-13\,$\mu$m wavelength range.
As a start, and for comparison purposes, we took a ring with zero inclination (face-on), so that the outer radius R$_{out}$ was the only free parameter.

Left panel of Fig.\,\ref{fig: face-on-incl-ring} shows the visiblities plotted as a function of {\it uv}-radius together with the best-fit model. 
For the latter, R$_{out}$=6.3\,AU\,$^{+0.1}_{-0.2}$ and the reduced $\chi^2$=2.4. Two conclusions can be drawn from this preliminary analysis. 
First, the mid-IR emission of R CrA is clearly resolved into a structure of $\sim$\,6\,AU radius. Second, this structure is asymmetric which hints at a 
inclined disk geometry. 
Including the inclination in the model leads to a better fit, with a reduced $\chi^2$=1.1. The best-fit model is shown in contour plot in 
the right panel of Fig.\,\ref{fig: face-on-incl-ring} together with the {\it uv}-coverage of our observations. The fitted parameters are R$_{out}$=8.6\,AU$^{+6.4}_{-1.9}$, 
an inclination of i=47$^\circ$\,$^{+21}_{-22}$ with respect to the plane of the sky, and a semimajor axis position angle\,\footnote{oriented east of north.} PA=152$^\circ$\,$^{+24}_{-14}$.

\subsection{Disk structure and orientation}
\label{sec: disk_orient_morphology}

\begin{figure}
\centering
\includegraphics[height=8.5cm]{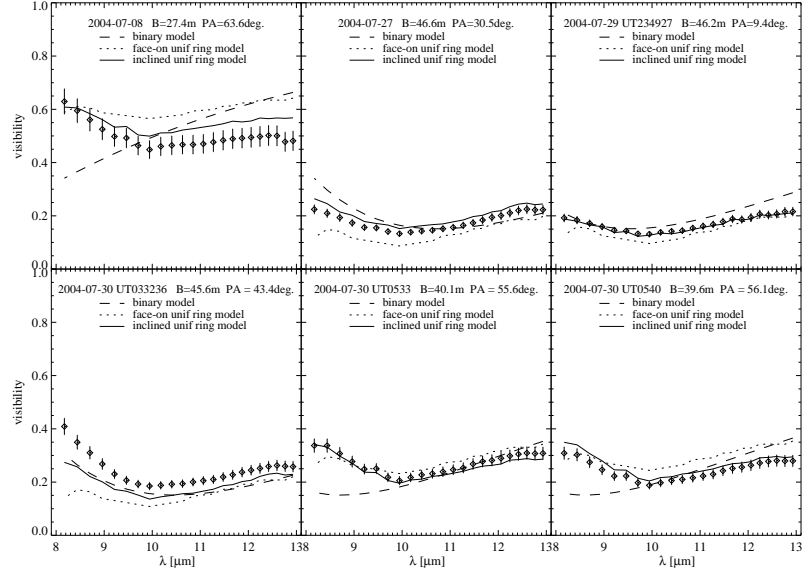}
\caption{Spectrally-dispersed visibilities between with 8-13\,$\mu$m (diamonds) with the best-fit models (binary, face-on and inclined uniform rings). }
\label{fig: spect-disp-vis}      
\end{figure}

\begin{figure}
\centering
\begin{tabular}{cc}
\includegraphics[height=7.3cm]{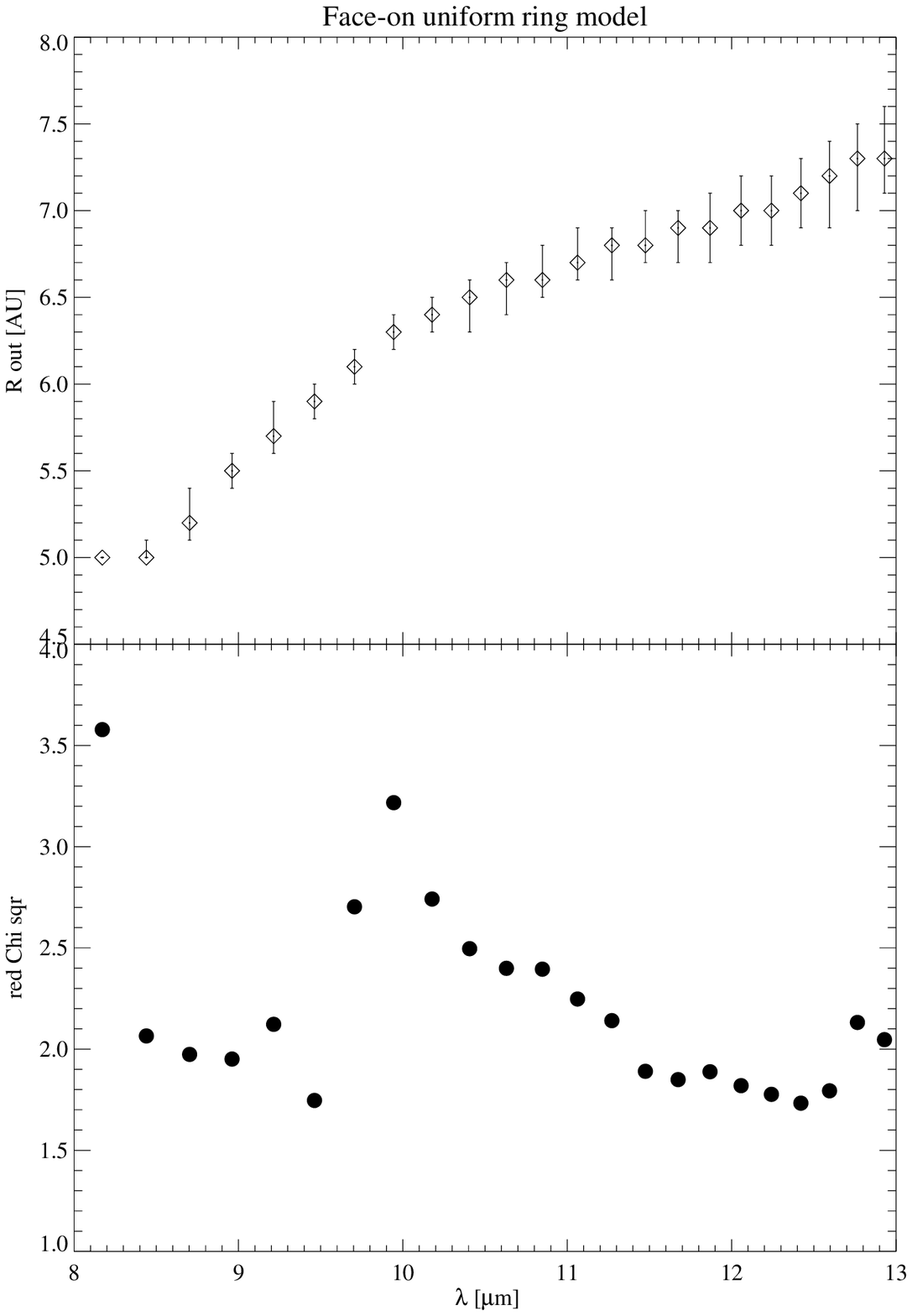} &
\includegraphics[height=7.3cm]{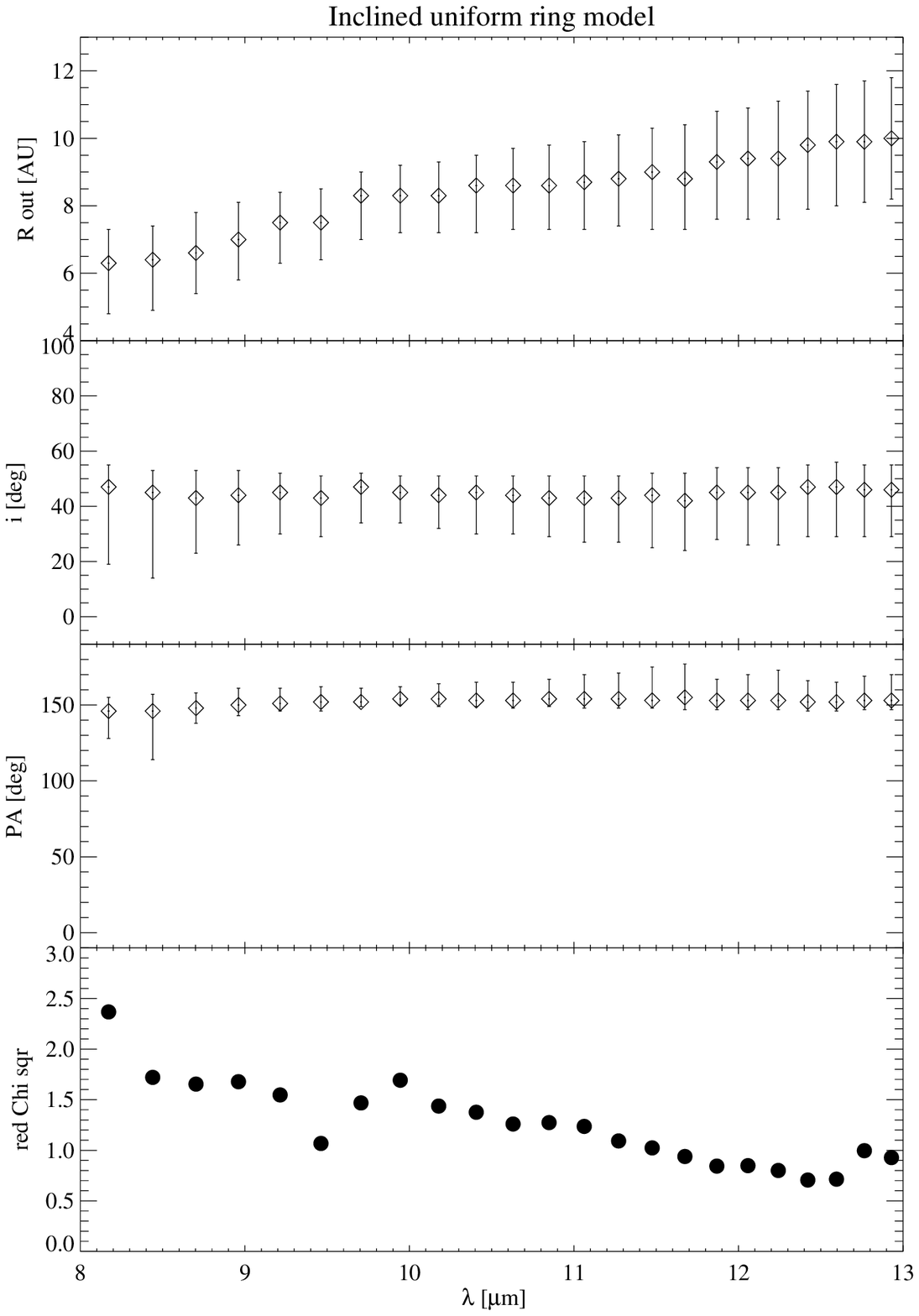}
\end{tabular}
\caption{Best fit parameters as a function of wavelength for respectively a face-on (left panel) and inclined (right panel) uniform ring model.}
\label{fig: wave_param}      
\end{figure}

Spectrally-dispersed visibilities as measured by MIDI are likely to provide additional information, because of the extension of {\it uv}-coverage. 
However, one should keep in mind that there is in principle a  degeneracy between spatial brightness distribution (morphology) and dust 
spectral emission features. A fortunate observation, in this respect, is that the ISO-SWS spectra are rather featureless apart from a broad 
9.7\,$\mu$m amorphous silicate feature, i.e. no prominent PAH feature at 8.6\,$\mu$m and no 11\,$\mu$m complex feature\,\cite{Acke_Van_Ancker_2004}. 
This allows us to derive meaningful results by fitting our simple uniform ring model to the observed spectrally-dispersed visibilities. 
Fig.\,\ref{fig: spect-disp-vis} shows the set of observed spectrally-dispersed visibilities together with the best-fit face-on and inclined ring models. 
Given the possibility that the observed visibilities may be due to the presence of a close companion\,\footnote{R CrA is flagged in 
Hipparcos as ''stochastic binary'' and ''suspected nonsingle''}, we additionally fit a binary model to the complete data set. 
The fitted parameters as a function of wavelength are presented in Fig.\,\ref{fig: wave_param}, while their wavelength-averaged values can be found in 
Table\,\ref{tab: bestfit_disks_par}. The quoted errors correspond to a variation of reduced $\chi^2$ of unity. 
While we can rule out with a high degree of confidence a binary model, the best agreement with these mid-IR visibility data is found for an inclined 
uniform ring model.  

Fitting a uniform ring model to the total data set for each spectral channel as we did leads to a monotonic variation of R$_{out}$ with wavelength 
(Fig.\,\ref{fig: wave_param}). Smaller outer radii are found at shorter wavelengths. This explains the fact that emission closer to the star arises 
from hotter dust and could in principle be used to derive the temperature profile of the circumstellar disk. 

\begin{table}
\centering
\caption{Best-fit models based on spectrally-dispersed visibilities. Uniform ring parameters are wavelength-averaged values. 
The binary model separation, PA and brightness ratio are quoted in the R$_{out}$, PA and i columns, respectively.
}
\label{tab: bestfit_disks_par}       
\begin{tabular}{lclll}
\hline\noalign{\smallskip}
		& \hspace{0.5cm} reduced $\chi^2$	        &	\hspace{0.5cm} R$_{out}$ (AU) 		&  \hspace{0.5cm} i ($^\circ$)		& \hspace{0.5cm} PA ($^\circ$)	 \\	
\noalign{\smallskip}\hline\noalign{\smallskip}
Binary			 & \hspace{0.5cm}2.4		&	\hspace{0.5cm} 3.1\,$^{+0.2}_{-0.2}$ 	& \hspace{0.5cm} 0.44\,$^{+0.05}_{-0.04}$ & \hspace{0.5cm}  36\,$^{+10}_{-10}$ 		\\
Face-on uniform ring & \hspace{0.5cm}2.2$\pm$0.7	 &	\hspace{0.5cm} 6.43\,$^{+0.15}_{-0.15}$ 	& \hspace{0.5cm} ...  		  	& \hspace{0.5cm} ...   		\\
Inclined uniform ring	 & \hspace{0.5cm}1.3$\pm$0.6 &	\hspace{0.5cm} 8.48\,$^{+1.26}_{-1.47}$ 	& \hspace{0.5cm} 44\,$^{+8}_{-17}$ & \hspace{0.5cm} 152\,$^{+14}_{-7}$    \\
\noalign{\smallskip}\hline
\end{tabular}
\end{table}

The inclination of the ring with respect to the plane of the sky is 44$^\circ$\,$^{+8}_{-17}$. This is fully consistent with the inclination of the plane (40$^\circ$) 
perpendicular to the symmetry axis of the bipolar reflection nebula as suggested from NIR imaging polarimetry\,\cite{Clark_etal_2000}. 
However, the derived position angle of the disk symmetry axis (62$^\circ$\,$^{+14}_{-7}$) is only marginally consistent with R CrA as the driving source 
for the extended NE-SW molecular outflow at PA$\sim$\,30$^\circ$\,\cite{Levreault_1988},\cite{Anderson_etal_1997}. The same conclusion holds 
for the direction of the closest HH objects with respect to R CrA\,\cite{Wang_2004} (HH\,96\,: $\sim$\,38$^\circ$, HH\,97W\,: $\sim$\,34$^\circ$, 
HH\,98\,: $\sim$\,33$^\circ$, HH\,100\,: $\sim$\,34$^\circ$, HH\,99A\,: $\sim$\,51$^\circ$, HH\,104A\,: $\sim$\,95$^\circ$).  






\vspace{0.3cm}
\noindent {\it Acknowledgment}\,: We are grateful to B. Wilking for his help. GM is supported by DFG grant ME 2061/3-1.





%
%
%
%
%
%
%
%
%
%
%
%
%
\input{correia_ref}



\printindex
\end{document}

%% file: correia_ref.tex
%
%

%
%